\documentclass[12pt]{iopart}
\usepackage {epsfig}% Include figure files
\usepackage {graphicx}% Include figure files
\begin{document}

\title[Role of transition metal impurities and oxygen vacancies]{The role 
of transition metal impurities and oxygen vacancies in the
formation of ferromagnetism in Co-doped TiO$_2$}

\author{V I Anisimov$^1$, M A Korotin$^1$, I A Nekrasov$^2$, A~S~Mylnikova$^{1,3}$, 
A V Lukoyanov$^{1,3}$, J L Wang$^4$ and Z Zeng$^4$}
\address{$^1$ Institute of Metal Physics, Russian Academy of 
Sciences, Ural Division, 620041 Ekaterinburg GSP-170, Russia}
\address{$^2$ Institute of Electrophysics, Russian Academy of 
Sciences, Ural Division, 620049 Ekaterinburg, Russia}
\address{$^3$ Ural State Technical University-UPI, 620002 Ekaterinburg, Russia}
\address{$^4$ Key Laboratory of Materials Physics, Institute of Solid 
State Physics, Chinese Academy of Sciences,
P.O. Box 1129, Hefei 230031, China}

\ead{via@optics.imp.uran.ru}
\begin{abstract}
We have investigated the role of transition metal impurities and oxygen
vacancies in the formation of ferromagnetism in Co-doped TiO$_2$ using
LSDA+U approach which takes into account strong on-cite Coulomb
correlations for electronic structure calculations. Several model systems
such as supercells of rutile TiO$_2$ with Co$^{+2}$ ion in high-spin state
substituted into titanium cite and with oxygen vacancies were considered.
We found that exchange interaction of Co magnetic ions is ferromagnetic,
but very weak due to the large average impurity-impurity distance. 
However, its strength becomes three times stronger when there is a magnetic
vacancy nearby. The magnetic moment values are 3~$\mu_B$ for Co$^{2+}$ and
1~$\mu_B$ for vacancy with the opposite directions of them. 
Our investigation showed that exchange interaction energy of Co 
and vacancy moments varies from 330~meV to 40~meV depending on the 
distance between them in the supercell. We suppose that strong interaction 
between Co and vacancy moments should be taken into account 
for the explanation of high Curie temperature value in Co-doped TiO$_2$.

\end{abstract}

\pacs{75.30.Hx, 71.55.Ht}
\submitto{\JPCM}

\maketitle

\section {Introduction}

Diluted magnetic semiconductors (DMS's) have been studied extensively
because of the potential possibility to use both charge and spin degrees of
freedom of carriers in the electronic devices, namely in the
spintronics~\cite {spintronics1, spintronics3}. However, the
most investigated II-VI and III-V compounds doped with magnetic transition
metal ions have Curie temperature (T$_C$) about 110~K or less. 

Recent research effort has been focused on developing new ferromagnetic
semiconductors operating at room temperature. In this connection oxide
semiconductors became a subject of intent interest~\cite {spintronics3, 
CIM2, ODMS.list1}. In 2001 Matsumoto {\it et al.}~\cite {Matsumoto}
reported that Co-doped anatase TiO$_2$ can keep ferromagnetic order up to
400~K with magnetic moment about 0.32~$\mu_B$/Co for Co concentration up to
8~\%. More recently, Park {\it et al.}~\cite {Park} have successfully grown
the ferromagnetic Co-doped rutile TiO$_2$ films. The Curie temperature was
estimated to be above 400~K for 12~\% of Co content. By now a lot of other
semiconductors showing room-temperature ferromagnetism have been
fabricated~\cite {spintronics3, CIM2, list2}.

Despite recent experimental success, there is no consistent theoretical 
description of DMS's. At first, ferromagnetism in Co-doped TiO$_2$ was
explained in terms of carrier induced mechanism like in III-V based
DMS's~\cite {1explTiO2}. Samples fabricated under oxygen-rich
conditions show negligible magnetization and are insulators \cite
{Properties1, CIM4.properties2} so that one can suppose that 
oxygen vacancies, which are easily formed in thin films, are essential to
ferromagnetic order and conductivity. The magnetic moment value also depends on
the sample growth conditions. For example, it is reported to be
1.26~$\mu_B$/Co in~\cite {Properties3} to compare with
0.32~$\mu_B$/Co, obtained by Matsumoto {\it et al.}~\cite {Matsumoto}. The
most recent measurements~\cite {Griffin05} yield a spontaneous
magnetization value of 1.1~$\mu_B$/Co atom for the films which were highly
insulating with Co in 2+ state.

The problem of the role of magnetic ions doping and oxygen deficiency in
the origin of room-temperature ferromagnetism was considered in many
theoretical papers. For example, in~\cite {Park.calc, H.Weng}
first-principles calculations were performed to consider the influence of
vacancy position on magnetic properties of Co-doped TiO$_2$, in 
particular, on Co spin state. Park {\it et al.}~\cite {Park.calc} treated 
an oxygen vacancy in the supercell (Ti$_{15}$Co$_1$O$_{31}$) by the LSDA+U method 
with spin-orbit coupling taken in spin-diagonal form. They showed that for the anatase 
structure of Ti$_{1-x}$Co$_x$O$_2$ (x = 0.0625 and 0.125) the presence 
of oxygen vacancy near Co results in the state with the
spin magnetic moment of 2.53~$\mu_B$, whereas the presence of vacancy near
Ti does not affect the magnetic moment value distinctly. However, according 
to their results, oxygen vacancy near Ti site is more stable. This conflicts 
with~\cite {H.Weng} where the opposite is claimed: the oxygen vacancy prefers 
to stay near Co, and it causes the magnetic moment on Co equal to 0.90~$\mu_B$. 
In~\cite {ODMS.list1} overlap polaron model caused by shallow donor electrons 
was proposed to explain high Curie temperatures in Co-doped TiO$_2$. 
Unlike super-exchange or double-exchange interaction, 
ferromagnetic exchange is suggested to produce long-range order. 
The other recent theoretical works are devoted to the role of interstitial Co
that might appear in the material~\cite {Geng, Sullivan03}. From one point
of view, interstitial Co destroys the spin polarization of the
substitutional Co nearby and, hence, reduces the average impurity magnetic
moment~\cite {Geng}. From the other results, there is an enhancement of the
average value of the local magnetic moment relative to the Co$^{2+}$ low
spin state with the presence of substitutional Co~\cite {Sullivan03}. Some
theoretical investigations claimed that Co impurities have strong tendency for
clusterization~\cite {Geng, Jaffe05}. Process of preparation and concentrations 
of Co impurity can lead to Co metal clusters and/or dispersed matrix-incorporated 
Co~\cite {Shinde}.

At the same time, the authors 
of~\cite {Matsumoto, Properties1, Properties3} confirm the absence
of Co metal clusters. Thus, there are many aspects of physical properties
of Co-doped TiO$_2$ which are under debate up to date.

The main purpose of this work was to clarify the role of Co impurities and
oxygen deficiency in the origin of ferromagnetism with high T$_C$ in
Co-doped TiO$_2$. Since tetravalent Ti$^{4+}$ ion is substituted by
divalent Co$^{2+}$, there are intrinsic oxygen vacancies in doped system
to compensate this charge difference. The actual value of oxygen deficiency
is difficult to estimate because defect concentration strongly depends on
the growth conditions. We have considered several model systems with
various content of Co impurity ions and vacancies. In our study
we assumed that there is no Co clustering, Co ions are substitutional and 
their valency state is +2. Our
investigations were performed using the linearized muffin-tin orbital
(LMTO) band structure calculation method within the local spin density
approximation with the on-site Coulomb repulsion $U$ (LSDA+U)~\cite
{tblmto, LDA.LDA+U}. The lattice relaxation was neglected.

\begin {figure}
\centering
\includegraphics [clip=true,width=0.9\columnwidth] {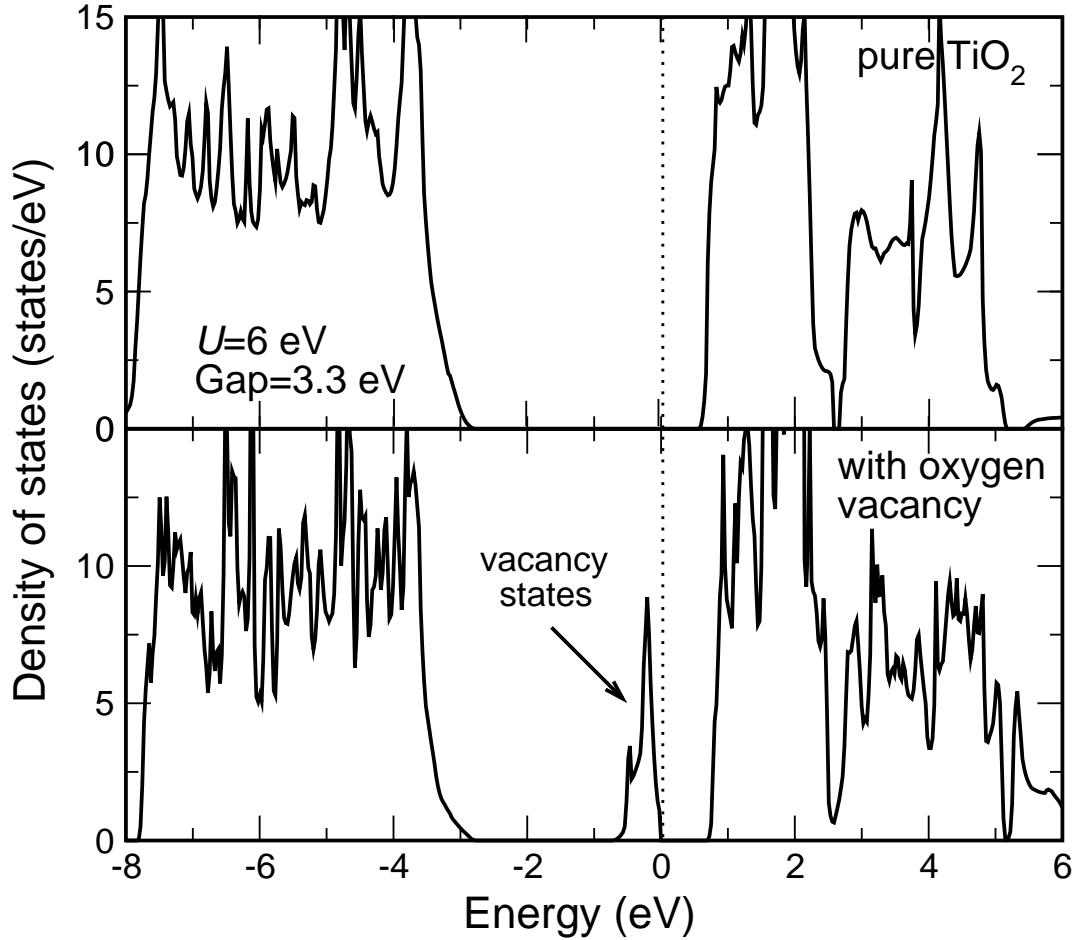}
\caption {Total densities of states obtained from the LSDA+U calculation
for pure TiO$_2$ (top panel) and model oxygen deficient TiO$_{2-1/4}$ (bottom
panel) compounds. The figures were aligned by the edge of oxygen bands. Zero 
of energy is at the top of the vacancy band for TiO$_{2-1/4}$.}
\label {Pure}
\end {figure}

The analysis was started with calculation of the electronic structure of
TiO$_2$ in stoichiometric case without doping and defects. TiO$_2$ has
three kinds of crystal structures: rutile, anatase and brookite. For our
calculations we choose the first one with crystallographic parameters,
taken from~\cite {structure}. The radii of MT spheres for LMTO
calculations were R$_{\rm Ti}$=2.42~a.u. and R$_{\rm O}$=1.85~a.u.; two
types of empty spheres (atomic spheres with zero nuclear charge) were
added. The valence band is formed by mainly oxygen 2$p$ orbitals
contribution, and the conduction band by Ti 3$d$ orbitals. 
The LDA calculation gave the value of band gap equal to 1.7~eV 
which is significantly less than the experimental value 
of 3.1~eV~\cite {Matsumoto}. Mo et al. ~\cite {Mo2002} shown that while LDA underestimates the energy gap, 
other ground state properties such as equilibrium lattice constants and bulk
moduli are in good agreement with experimental data. It is well-known problem of band gap
underestimation in LDA approach, where potential is orbital-independent. It
can be improved by means of LSDA+U method, where orbital-dependent
potential acts in different way on the occupied and unoccupied
$d$-orbitals~\cite {LDA.LDA+U}. In the top panel of figure~\ref {Pure} the
electronic structure obtained from the LSDA+U calculation is shown. The
values of on-site Coulomb interaction parameters $U$ and $J_H$ for Ti 3$d$
orbitals were chosen to be 6.0~eV and 0.7~eV, respectively~\cite
{LDA.LDA+U}. The calculated band gap value is 3.3~eV in good agreement 
with the experimental value. The LMTO radii and U and $J_H$ values were taken to
provide experimental gap value, they weakly influence the magnetic properties. 

Below we present results of our LSDA+U calculations for model systems
corresponding to 1) Co impurities in TiO$_2$ without oxygen deficiency
(section \ref {Co-doped TiO2}), 2) oxygen deficient TiO$_2$ but without Co
impurities (section \ref {vacancy}) and 3) the simultaneous presence of Co
impurities and oxygen vacancies in TiO$_2$ (section \ref {Co-V}). 

The oxygen vacancies play dual role in this material. In the first, every
vacancy adds two electrons to the system and in the second, it produces
localized states in the energy gap. In order to separate the influence of 
those factors on the electronic structure and magnetic state of the system
we imitated the first one in the calculation by adding electrons to the supercell 
with compensating their negative charge by the uniform charge of the opposite
sign distributed over the supercell. This has a meaning of taking into
account the presence of the vacancy in the crystal but somewhere far away
from the atoms considered in the model calculations so that one can neglect
the influence of the localized states produced by this vacancy. 

The substitution of tetravalent Ti$^{4+}$ ion by divalent Co$^{2+}$ has an
effect of adding two holes to the system. This could be also imitated in
our calculations by removing two electrons from the supercell with the adding
the corresponding uniform charge of the opposite sign. This adding
(removing) of the electrons is not simply shift of Fermi energy level as in
the ``rigid band'' approximation because we performed fully self-consistent
calculations with the changed number of electrons and took into account the
change of the charge density distribution and hence electron potential.

In the present study we adjusted total number of
electrons in the system so that Co ions were always in divalent state
(experimental works report \cite {Properties1,Properties3,CoValency3} that
all Co ions are in +2 formal oxidation state). In each model calculation with Co impurity the supercell of
TiO$_2$ was considered to make impurity concentration close to the experimental
values of 8-12~\%. For vacancy states we have considered all three possible configurations: unoccupied,
occupied with one electron and fully occupied with two electrons.

\section {Cobalt-doped titanium dioxide}
\label {Co-doped TiO2}

To investigate the electronic structure of Co-doped TiO$_2$, we used a
supercell containing sixteen TiO$_2$ formula units with two Ti replaced by
Co where impurities are separated by 6.59~\AA. Such supercell corresponds 
to 12.5~\% of Co content. 

\begin {figure}
\centering
\includegraphics [clip=true,width=0.9\columnwidth] {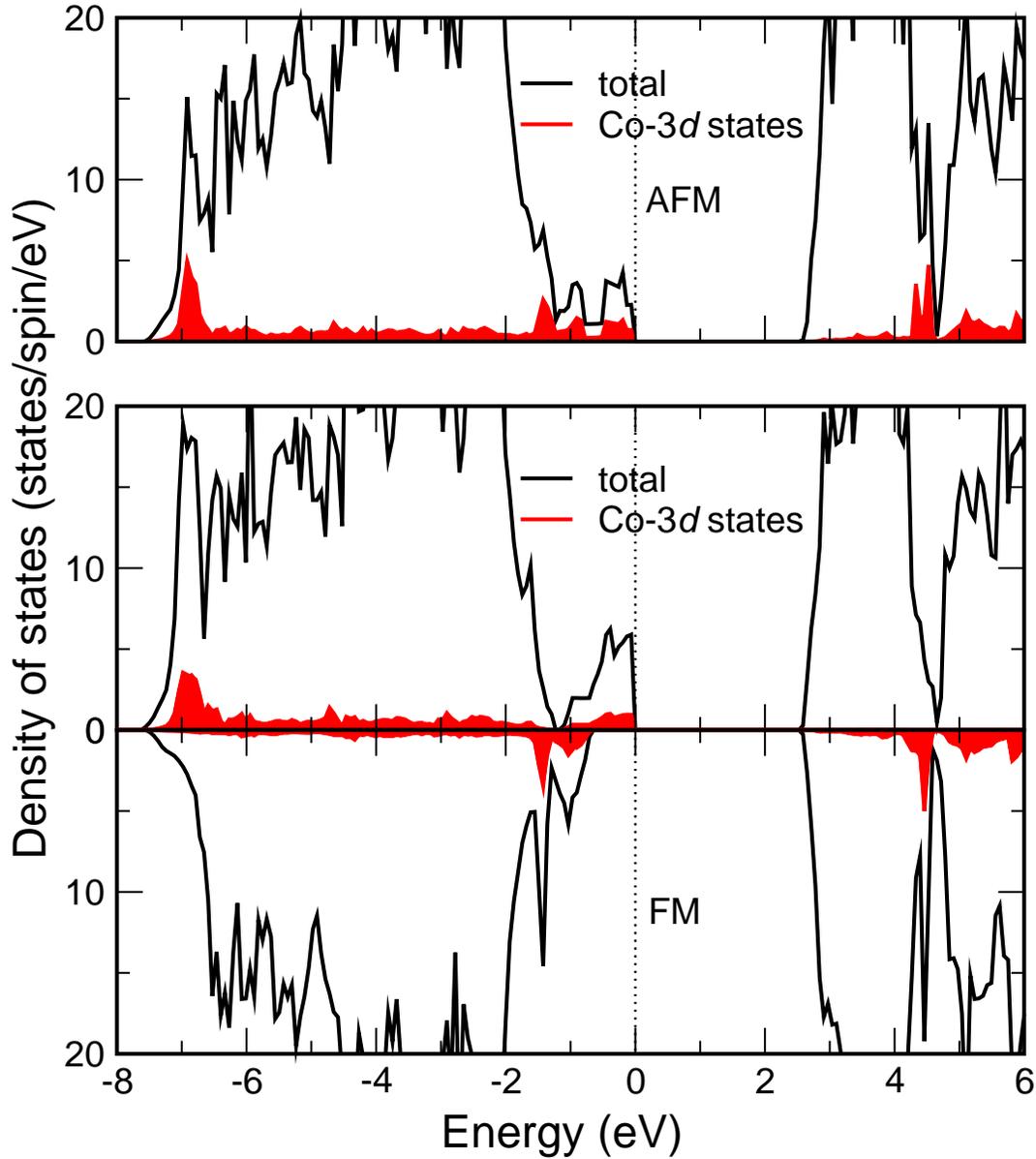}
\caption {Total and partial densities of states obtained
from the LSDA+U calculations for antiferromagnetic (AFM; top panel) and
ferromagnetic (FM; bottom panel) configurations of Co ions in 
Co$_{1/8}$Ti$_{1-1/8}$O$_{2}$ model compound.}
\label {2Co201}
\end {figure}

In order to reproduce Co ions in +2 formal oxidation state Co$^{2+}$
(according to experimental evidence \cite {Properties1, Properties3,
CoValency3}) we have added to the system two electrons per Co ion (four
electrons per supercell). Physically that is equivalent to considering the
system with one oxygen vacancy per Co ion
(Co$_{1/8}$Ti$_{1-1/8}$O$_{2-1/8}$ instead of Co$_{1/8}$Ti$_{1-1/8}$O$_{2}$
model compound) but without explicitly taking into account the localized
states produced by the vacancy in the energy gap.

The resulting density of states has an insulating character. In figure~\ref
{2Co201} we present the electronic structure for 
Co$_{1/8}$Ti$_{1-1/8}$O$_{2}$ model compound obtained from the LSDA+U band
calculations. The values of on-site Coulomb interaction parameters $U$ and
$J_H$ for Co 3$d$ orbitals were chosen as 8~eV and 1~eV,
respectively~\cite {LDA.LDA+U}.

Because of large distances between impurities due to small Co
concentration, one can expect that exchange interaction between Co ions is
weak. To estimate the exchange interaction parameter, we examined the
total energy difference between ferromagnetic (FM) and antiferromagnetic
(AFM) orientations of Co ions magnetic moments. Our results showed that FM
phase is lower in energy than AFM by 3~meV (or $\sim$30 K; that is an
order of magnitude smaller then experimental value of T$_C$). The obtained
magnetic moments values for Co ions in the ground state are 3~$\mu_B$.
Thus, exchange interaction of magnetic ions is FM, but weak even for the
impurity concentration larger than ones in fabricated samples of Co-doped
TiO$_2$~\cite {Park}. Therefore exchange interaction between impurities magnetic
moments alone without taking into account other states cannot yield as high
T$_C$ as observed. 

\section {Titanium dioxide with oxygen vacancies}
\label {vacancy}

In order to explore the electronic structure and magnetic properties
changes caused by the localized states produced by oxygen vacancies, we
at first considered a supercell containing four formula units of TiO$_2$ with
one oxygen atom replaced by an empty sphere; it corresponds to
TiO$_{2-1/8}$ composition. Result of the LSDA+U calculation is shown in
figure~\ref {Pure} (bottom panel). Electronic structure for oxygen vacancy
exhibits narrow defect band, which is fully occupied. It is formed of
mainly Ti $d$-states of the Ti ions nearby the vacancy split off from the
bottom of conduction band and has a symmetry of $s$ orbital centered on the
vacancy site. Two electrons added to the system as the result of removing
O$^{-2}$ ion occupy this band of $s$ symmetry and result in a nonmagnetic
ground state. To obtain a magnetic solution and, consequently, localized
moment, this band should be partially filled. That corresponds to a
so-called $F^{+}$-center with one electron localized on an oxygen vacancy state.

To study half-filled vacancy band (still without explicitly considering
cobalt impurity), we added one hole per supercell. Physically such
occupancy of the defect states could be realized if the system contains one
Co$^{2+}$ impurity per two oxygen vacancies or one Ti vacancy per four
oxygen vacancies. In all our further calculations we added the number of
holes which would provide half-filled band formed of oxygen vacancies.

\begin {figure}
\centering
\includegraphics [clip=true,width=0.9\columnwidth] {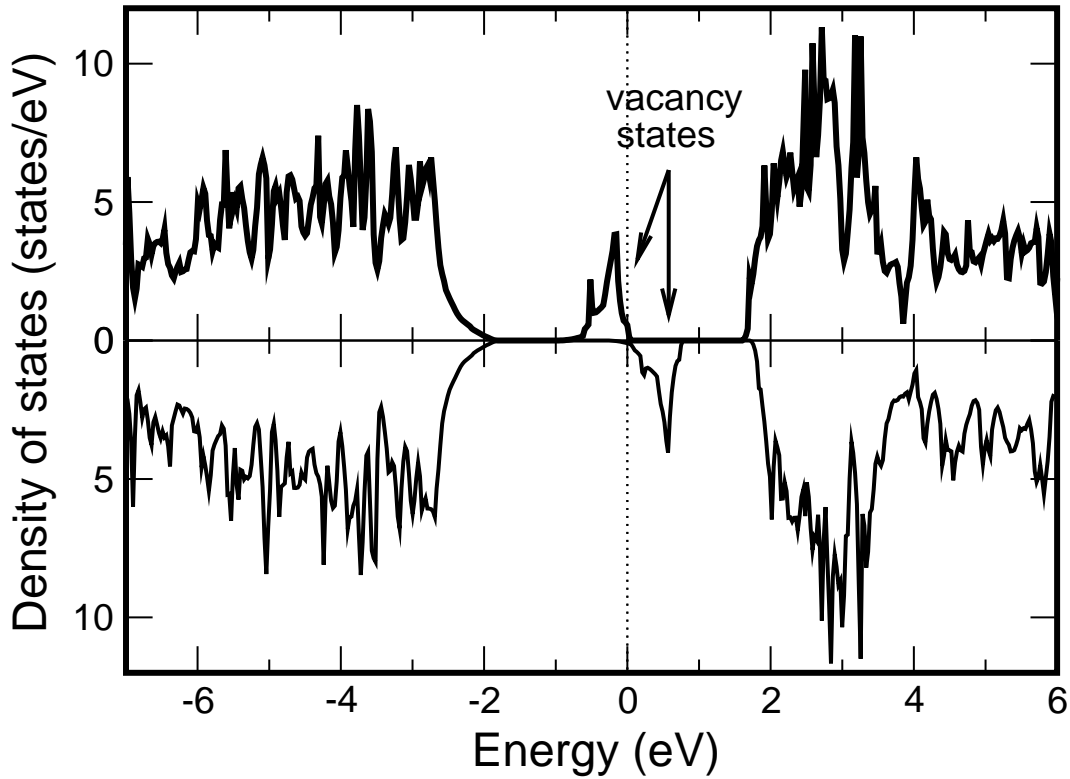}
\caption {Total density of states for oxygen deficient 
TiO$_{2-1/8}$ model compound in magnetic state. Fermi energy is zero.}
\label {defect_band}
\end {figure}

The partial occupancy of the defect band results in a spin-polarized
magnetic solution. The magnetic state has lower total energy in comparison
with nonmagnetic one; thus, it is the ground state of the system with the
magnetic moment value per supercell of 1~$\mu_B$. In figure~\ref
{defect_band} the density of states for this magnetic solution is presented.

Considering our result one can suppose that not only Co ions, but also
vacancies yield sizeable magnetic signal in Co-doped TiO$_2$. The similar
idea was proposed in recent paper~\cite {ZnO}, where observed
room-temperature ferromagnetism in ZnO was interpreted in terms of a 
spin-split donor impurity-band model, and $F^{+}$ centers, associated with
oxygen vacancies, were considered as a possible source of magnetism. In
principle, vacancies themselves without magnetic impurities could induce
magnetization as was shown in the work of Monnier {\it et al.}~\cite
{Monnier}. They observed high temperature ferromagnetism in normally pure 
CaB$_6$, fabricated under Ca-poor growth conditions. In these crystals 
strong covalency of the B-B bonds makes B-vacancies more energetically 
preferable to form. Large values of magnetic moments could be obtained 
in the distorted lattices where vacancies are located nearby.

We have also investigated exchange interaction between vacancies,
considered FM and AFM configurations of vacancies moments. We dealt with a
supercell containing eight formula units with two oxygen vacancies in a
supercell; it corresponds again to TiO$_{2-1/8}$ composition. The results
of the LSDA+U calculations showed that AFM solution is lower in energy by
50~meV. That can be explained by the Hubbard model theory. It predicts
that AFM solution is the ground state in the case of half-filled band. 

The next step is to study how vacancy moment interacts with Co one and how 
the presence of vacancy influences on Co-Co interaction. 

\section {Cobalt-doped titanium dioxide with oxygen vacancies}
\label {Co-V}

In order to explore the influence of vacancy states on the magnetism of
Co-doped TiO$_{2}$, we considered a supercell containing eight formula
units with one Co impurity and one oxygen vacancy at three different
possible distances from each other. The energy difference between FM and
AFM configurations of Co and vacancy was calculated for each case within
LSDA+U approach to estimate the strength of exchange interaction between Co
and vacancy magnetic moments. As in the previous section, total number of
electrons in the supercell was adjusted to produce divalent Co$^{2+}$ ions
and half-filled oxygen vacancy states ($F^{+}$-centers). That required adding
one electron to the supercell because every vacancy adds two electrons to
the system and two electrons are removed when tetravalent Ti$^{4+}$ ion is
substituted by divalent Co$^{2+}$. That could be realized by the presence
of additional oxygen vacancy per Co-vacancy pair but without considering
explicitly the corresponding localized states. 

\begin {figure}
\centering
\includegraphics [clip=true,width=0.78\columnwidth] {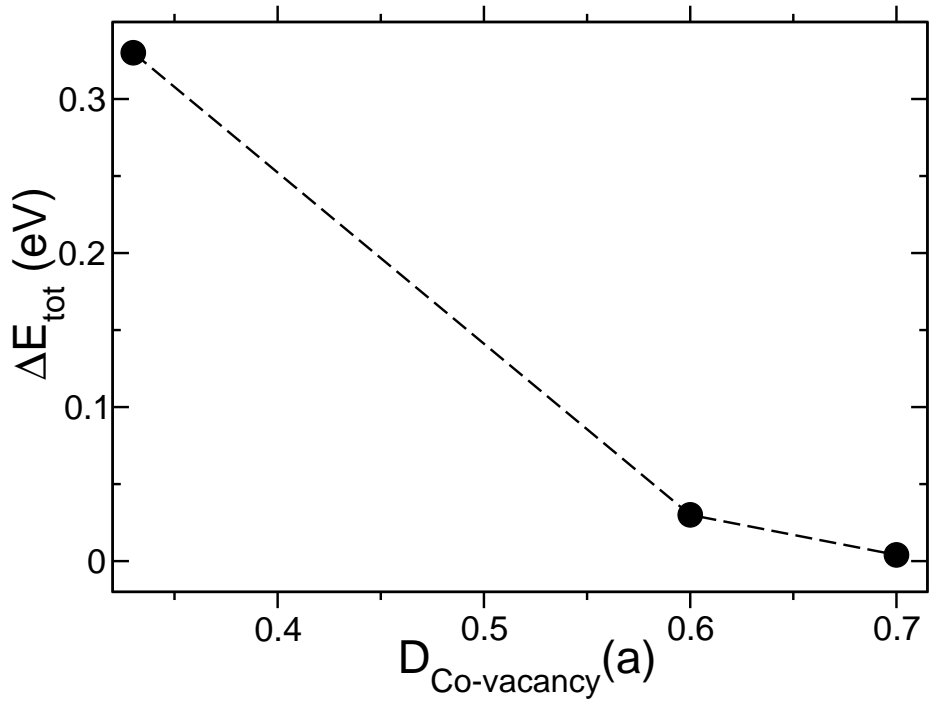}
\caption {The energy difference between FM and AFM configurations of Co and
vacancy moments as a function of distance between them for
Co$_{1/8}$Ti$_{1-1/8}$O$_{2-1/8}$ model compound. Distance is 
given in the units of lattice constant a=5.92~\AA.}
\label {DOTES}
\end {figure}

The obtained results are presented in figure~\ref {DOTES}. In all three
cases, AFM arrangement of Co and vacancy spins is lower in energy. The
calculated moments of Co and vacancy are 3~$\mu_B$ and -1~$\mu_B$,
correspondingly. Thus, oxygen vacancy prefers to have the magnetic moment
that is AFM ordered to Co moment. One can see (figure~\ref {DOTES}) that
strength of interaction between Co and vacancy drops rapidly with the
increase of the distance between them. For the nearest and the longest
distances of 1.9~\AA \ and 4.1~\AA, the energy difference values are
330~meV and 40~meV, correspondingly. This energy interval corresponds to
the temperature range from $\sim$3800~K to 500~K. We suppose that the
presence of strong exchange interaction between Co and vacancy moments
might results in a strongly coupled system of Co and vacancies magnetic
moments with a high ordering temperature.

\begin {figure}
\centering
\includegraphics [clip=true,width=0.9\columnwidth] {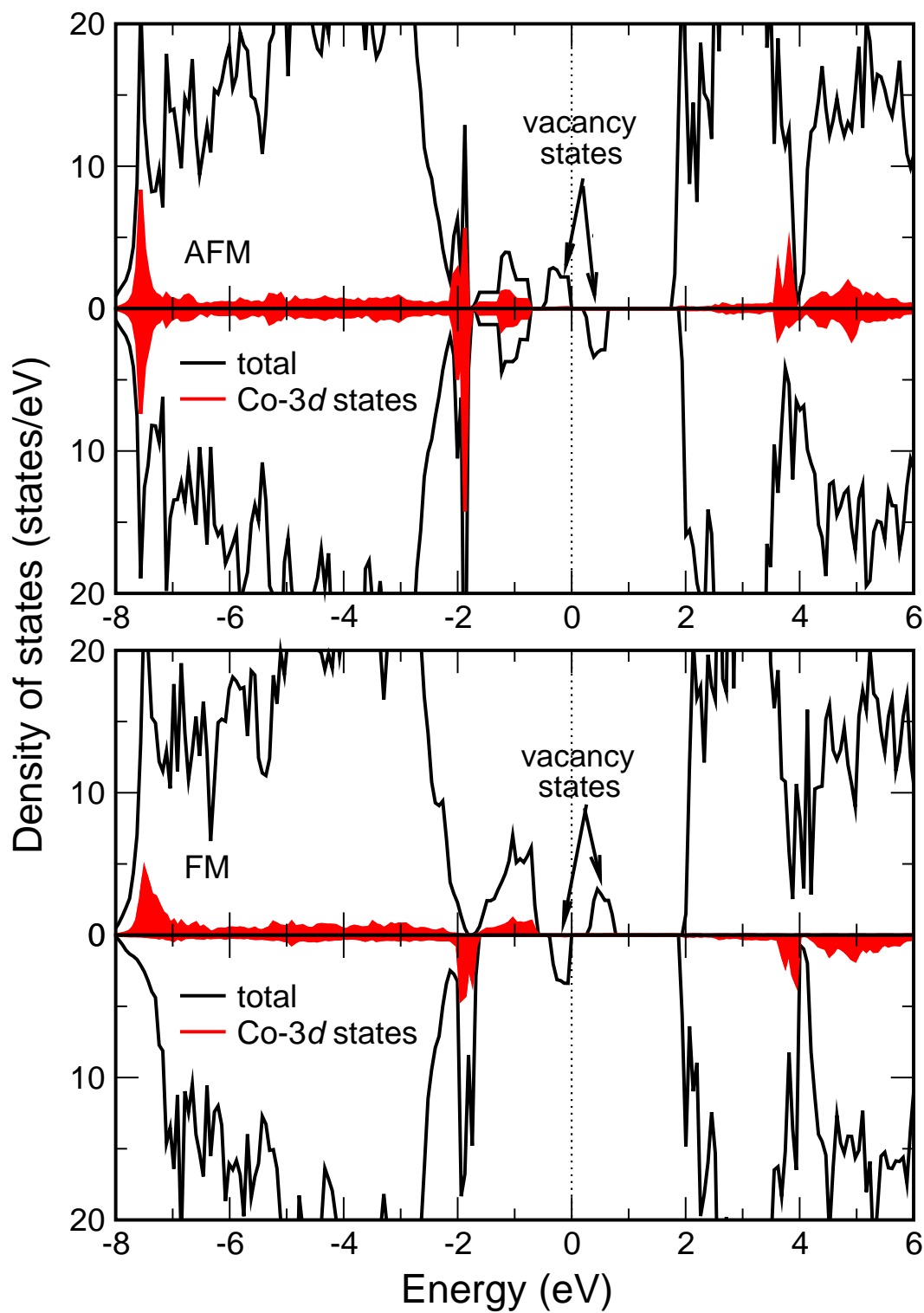}
\caption {Total and partial densities of states for AFM 
(top panel) and FM (bottom panel) configurations of Co ions in
Co$_{1/8}$Ti$_{1-1/8}$O$_{2-1/16}$ model compound obtained from the LSDA+U
calculations. Zero of energy is at the Fermi energy. }
\label {FIG5}
\end {figure}

To investigate vacancy influence on the strength of exchange interaction
between Co spins, we used the same supercell as in section~\ref {Co-doped
TiO2}, but with one oxygen ion removed from it. We chose the position of
the oxygen vacancy which is at the equal distance of 3.58~\AA \ from both
Co ions. Again total number of electrons in the supercell was adjusted to
produce divalent Co$^{2+}$ ions and half-filled oxygen vacancy states
($F^{+}$-centers). In this case that required adding three electron to the
supercell. (Similarly to the previous case that could be realized by the
presence of three additional oxygen vacancies per supercell but without
considering explicitly the corresponding localized states.)

Total and partial densities of states for AFM and FM configurations of Co 
ions obtained from the LSDA+U calculations are shown in figure~\ref {FIG5}. 
One can see that vacancy induced states are situated around the Fermi level 
between occupied and unoccupied Co 3$d$ states without overlapping them. 
This electronic structure rather agrees with the scheme proposed for Mn impurity 
in a spin-split donor impurity-band model,~\cite {ZnO} where there is no overlap 
of vacancy band with 3$d$ states. But in contrast with that, vacancy states 
are not in the middle between Co majority and minority states, but shifted closer 
to the occupied 3$d$ states, although without overlapping, like in the scheme 
proposed for Ti in ~\cite {ZnO}.

From the calculations we found that strength of magnetic ions interaction
in the presence of vacancy is three times stronger than that without
vacancy. The corresponding energy difference between AFM and FM
configurations of Co ions is equal to 10~meV; FM orientation of Co magnetic
moments is lower in energy. The obtained total cell magnetization is
5~$\mu_B$. At that each Co accounts for 3~$\mu_B$. Vacancy moment is equal
to 1~$\mu_B$ and is AFM ordered with respect to Co moments. The solution
with FM order of moments is not stable. Again, oxygen vacancy moment
prefers to be AFM ordered to Co one like in the previous case of one Co and
vacancy in the cell, which effectively results in a FM coupling between Co
ion moments.

\begin{table}
\caption{The list of supercells used in calculations: model of the calculation, 
type of magnetic ordering between Co ions, oxygen vacancies (Vac) or Co-vacancy, 
$\Delta$ E is the difference between FM and AFM configurations. 
In the third and fourth models the distance (D) between Co and vacancy is given. 
Last two cases correspond to half-filled and empty/full-filled vacancy states.}
\begin{indented}
\item[]\begin{tabular}{lll}
\br
Model & Type of ordering & $\Delta$ E, meV \\
\mr
Co$_{1/8}$Ti$_{1-1/8}$O$_{2-1/8}$ (16 f.u., 2 Co)&Co-Co&3\\
TiO$_{2-1/8}$ ~~~~~~~~~~~~~~( 8 f.u., 2 Vac)&Vac-Vac&50\\
Co$_{1/8}$Ti$_{1-1/8}$O$_{2-1/8}$ ( 8 f.u., 1 Co, 1 Vac) &
Co-Vac (D = 1.9~\AA)&330\\
Co$_{1/8}$Ti$_{1-1/8}$O$_{2-1/8}$ ( 8 f.u., 1 Co, 1 Vac) & Co-Vac (D
= 4.1~\AA)&40\\
Co$_{1/8}$Ti$_{1-1/8}$O$_{2-1/8}$ (16 f.u., 2 Co, 1 Vac)&Co-Co~ (half-filled)&10\\
Co$_{1/8}$Ti$_{1-1/8}$O$_{2-1/8}$ (16 f.u., 2 Co, 1 Vac)&Co-Co~
(empty/full-filled)&3\\
\br
\end{tabular}
\end{indented}
\end{table}

The next step in the investigation of vacancy influence on Co spins
interaction is to examine it when defect state band is empty or full, and,
hence, vacancy is nonmagnetic. As vacancy states are situated between Co
majority and minority $d$-states (figure~\ref {FIG5}), addition of an
electron or a hole makes the band full or empty, correspondingly, but does
not change Co valence. Energy difference between FM and AFM configurations
of Co ions appears to be about 3~meV in the case of both empty and full
defect state band, like in the supercell without vacancy. Thus, the
presence of oxygen vacancy enhances Co ions interaction only when the
vacancy is magnetic.

\section {Conclusion}

In summary, using LSDA+U approach we have investigated the role of magnetic
impurities and oxygen vacancies in the formation of ferromagnetism with
high T$_C$ in Co-doped TiO$_2$. We found that exchange interaction of
magnetic Co ions is ferromagnetic, but very weak due to the large average
impurity-impurity distance. However, its strength becomes three times
stronger when there is a magnetic vacancy present. The obtained magnetic
moments are 3~$\mu_B$ and 1~$\mu_B$ for Co and vacancy, respectively. We
also found that interaction between Co and vacancy moments is surprisingly
very strong even for the longest distance of 4.1~\AA \ in the supercell. In
this case energy difference between FM and AFM configurations of Co and
vacancy moments is 40~meV, that corresponds to $\sim$500~K. Our results
lead us to think that the observed magnetic signal cannot be attributed to
Co ions only: vacancies moments should be taken into consideration too. It
seems that strong interaction between Co and vacancy moments is a key
moment for explanation of high T$_C$ in Co-doped TiO$_2$.

\ack

The authors are very grateful to D. Vollhardt for the helpful discussions
and to V.V. Mazurenko for the critical reading of the manuscript. This
work was supported by Russian Foundation for Basic Research under the
grants RFFI-04-02-16096 and RFFI-03-02-39024, by the joint UrO-SO Project
N~25, Netherlands Organization for Scientific Research through NWO
047.016.005, RFBR 05-02-16301 (IN), programs of the Presidium of the
Russian Academy of Sciences (RAS) ``Quantum macrophysics'' and of the
Division of Physical Sciences of the RAS ``Strongly correlated electrons in
semiconductors, metals, superconductors and magnetic materials''. I.A.N.
acknowledges support from the Dynasty Foundation and International Centre
for Fundamental Physics in Moscow program for young scientists 2005,
Russian Science Support Foundation program for young PhD of RAS 2005. 
A.V.L. acknowledges support from the Dynasty Foundation and International Centre
for Fundamental Physics in Moscow.

\end {document}